\documentclass[12pt]{article}

\usepackage{bm}
\usepackage[sort&compress, numbers, round]{natbib}
\usepackage{epsfig}
\usepackage{amsfonts}
\usepackage{color}
\usepackage{times}

\usepackage{graphicx}

\newcommand{\ben}{\begin{enumerate}}
\newcommand{\een}{\end{enumerate}}
\newcommand{\beq}{\begin{equation}}
\newcommand{\eeq}{\end{equation}}
\newcommand{\beqa}{\begin{eqnarray}}
\newcommand{\eeqa}{\end{eqnarray}}
\newcommand{\beqas}{\begin{eqnarray*}}
\newcommand{\eeqas}{\end{eqnarray*}}
\newcommand{\bit}{\begin{itemize}}
\newcommand{\eit}{\end{itemize}}
\newcommand{\ba}{\begin{array}}
\newcommand{\ea}{\end{array}}
\newcommand{\bd}{\begin{displaymath}}
\newcommand{\ed}{\end{displaymath}}
\newcommand{\bc}{\begin{center}}
\newcommand{\ec}{\end{center}}
\newcommand{\ve}{\varepsilon}

\newcommand{\xx}{{\bm x}}
\newcommand{\xxc}{{\bm X}_{c}}

\title{\bf Heterogeneous Cell Population Dynamics: \\
Equation-Free Uncertainty Quantification Computations}

\author{Katherine A. Bold, Yu Zou, and Ioannis G. Kevrekidis\thanks{{\tt To whom correspondence should be addressed: yannis@Princeton.edu;
+1-609-258-2818}} \\
Department of Chemical Engineering and \\
Program in Applied and Computational Mathematics \\
Princeton University \\
Princeton, NJ 08544 \\
USA \\*[0.1in]
Michael A. Henson \\
Department of Chemical Engineering \\
University of Massachusetts \\
Amherst, MA 01003 \\
USA}

\date{{\em Submitted to Biophysical Journal \\*[0.025in]
\today}}

\begin{document}

\maketitle

\bc
\section*{ABSTRACT}
\ec

\noindent
We propose a computational approach to modeling the
collective dynamics of populations of coupled heterogeneous
biological oscillators.
In contrast to Monte Carlo simulation, this approach utilizes
generalized Polynomial Chaos (gPC) to represent random properties of
the population, thus reducing the dynamics of ensembles of
oscillators to dynamics of their (typically significantly fewer)
representative gPC coefficients.
Equation-Free (EF) methods are employed to efficiently evolve these
gPC coefficients in time and compute their coarse-grained stationary
state and/or limit cycle solutions, circumventing the derivation of
explicit, closed-form evolution equations.
Ensemble realizations of the oscillators and their statistics can be
readily reconstructed from these gPC coefficients.
We apply this methodology to the synchronization of yeast
glycolytic oscillators coupled by the membrane exchange of an
intracellular metabolite. 
The heterogeneity consists of a single
random parameter, which accounts for glucose influx into a cell,
with a Gaussian distribution over the population.
Coarse projective integration is used to accelerate the evolution of
the population statistics in time.
Coarse fixed-point algorithms in conjunction with a Poincar\'e
return map are used to compute oscillatory solutions for the cell
population and to quantify their stability.
 \vspace{.5in}

\section*{INTRODUCTION}

Autonomous oscillations are observed in biological systems ranging
in complexity from microorganisms to human
beings~\citep{winfree:2001}.
Often these oscillations are generated
at the cellular level through positive feedback loops embedded in
gene regulatory or metabolic networks~\citep{goldbeter:1996}.
Robust strategies have evolved, based on intracellular coupling of
{\em multiple} oscillators (rather than on a single cellular
oscillator subject to degradation and failure).
A fundamental feature of such multicellular systems is the
synchronization of individual oscillators to produce a coherent
overall rhythm~\citep{strogatz:2003}.
Synchronized oscillators are responsible for rhythm generation at
second (heartbeat generation), daily (circadian timekeeping) and
monthly (menstrual cycles) time scales.
Malfunctioning of these interconnected oscillators can produce
disastrous consequences such as sudden heart attacks and epileptic
seizures.

The development of mathematical models to study the synchronization
of coupled oscillators has a long and beautiful
history~\citep{winfree:2001}.
Seminal contributions to the fundamental understanding of
synchronization have been made by rigorous mathematical analysis of
simple model systems~\citep{acebron:2005,strogatz:2000}.
The study of mechanistic models of multicellular biological systems
typically involves computational approaches such as dynamic
simulation and numerical bifurcation analysis.
Difficulties in applying such scientific computing tools are largely
determined by population model complexity, which in turn is
determined by the complexity of the individual cell model and the
number of cells included in the population.
Certain applications require both a detailed single cell
model and a large ensemble of single cells for meaningful
computational study.
For instance, a single mammalian circadian oscillator
contains multiple interconnected feedforward and feedback loops that
have been modeled with up to 73 coupled differential
equations~\citep{forger:peshkin:2003}.
Meanwhile, the mammalian circadian system is comprised of
approximately 10,000 individual oscillators that communicate via
neurotransmitter mediated coupling~\citep{michel:colwell:2001}.
The development of stochastic simulations to generate meaningful
population statistics is possible only if the model ensemble size is
sufficiently large.
Therefore, there is considerable motivation to develop
efficient simulation and bifurcation 
analysis techniques for large, heterogeneous
ensembles of coupled, complex biological oscillators.
Previous studies of heterogeneity 
\citep{Moon1:06,Moon2:06} used
idealized phase models (a single ``phase" equation
for each oscillator);
in this work, we show how the approach can 
be applied to ensembles of realistic limit cycle
oscillator models.

Among traditional techniques employed in the study of ensemble
statistics of stochastic systems, the most popular one is Monte
Carlo simulation \citep{Fishman:96} (or the cell-ensemble method
\citep{Schuler:83} in the biological context).
This approach, however, becomes extremely time-consuming when the
number of simulation realizations is large, as in the case of
multicellular coupled oscillator populations.
As an alternative, the stochastic
Galerkin method for uncertainty quantification (UQ) has been widely
used in recent years for solving stochastic ODEs or PDEs.
Pioneering work along these lines \citep{Ghanem:91} studied
stochastic systems with Gaussian random variables: by viewing
randomness as an additional dimension, beyond the space and time
dimensions, the dependence of system responses on random parameters is
represented in terms of orthogonal polynomial expansions of random
variables.
In this context (see Appendix A) all orthogonal polynomials of a
given order are called Polynomial Chaoses or Homogeneous Chaoses of
that order; projections of the system responses onto the Polynomial
Chaos (PC) coefficients evolve deterministically, and equations
for their evolution can in principle be obtained, and then solved,
by applying a Galerkin projection.
The method has been applied for uncertainty quantification purposes
in various physical and engineering systems including structures
with random properties \citep{Ghanem:91}, porous media
\citep{GhanemA:98}, fluid dynamics \citep{Maitre:01} and chemical
reactions \citep{Reagan:04}.
In \citep{XiuB:02}, the method was extended: {\em  generalized}
Polynomial Chaos (gPC), applicable to a variety of continuous and
discrete probability measures, was proposed based on the Askey
scheme.
In addition to representations of orthogonal polynomials, this
method was also improved in other directions, in the form of
piecewise (h-refinement) representations \citep{Deb:01} and wavelet
expansions \citep{Maitre:04}.

One advantage of the stochastic Galerkin method, as compared to
direct Monte Carlo simulation, is that it can reduce a stochastic
system to a deterministic one with (often significantly) fewer
degrees of freedom, thus accelerating computation and saving data
storage space.
In order to apply the stochastic Galerkin method, however, one must
derive equations for the temporal evolution of gPC coefficients
either explicitly or through a pseudospectral approach (e.g.
\citep{Reagan:03}) and develop a new code for the solution of these
equations.
To circumvent this additional effort, Equation-Free (EF) methods
\citep{Theodoropoulos:00,Kevrekidis:03,Kevrekidis:04} have been
utilized recently to quantify propagation of uncertainty in a
stochastic system by evolving gPC coefficients of random solutions
using the system dynamic simulator in a {\em nonintrusive way}, that
is, without deriving the corresponding explicit gPC evolution
equations \citep{Xiu:05}.
Within this multiscale  equation-free framework, the original stochastic
dynamics code is viewed as a fine-level simulator, while the
(unavailable) ODEs for the time-evolution of the gPC coefficients
are viewed as a coarse-grained system model.
These equation-free algorithms are built based on protocols that enable
communication between
different levels of system description; the {\em lifting} protocol
translates coarse-grained
initial conditions to one or more consistent fine scale initial
conditions; the {\em restriction}
protocol computes the coarse-grained description (values of the coarse
variables, ``observables")
of fine scale system configurations.
The success of this class of methods relies on the assumption that
closed evolution equations for the dominant (low-order) gPC
coefficients exist {\em in principle}, even if they are not
explicitly available.

The purpose of this paper is to demonstrate that efficient
simulation strategies for large populations of coupled biological
oscillators can be developed by utilizing these equation-free
uncertainty quantification (EF-UQ) based methods.
A six-dimensional cellular model of yeast glycolytic
oscillations~\citep{henson:muller:reuss:2002, Wolf:00} is
used to study synchronization of 1,000 {\em heterogeneous} cells in
a well mixed environment.
The random variable (which characterizes the cell population
heterogeneity) is chosen as the glucose influx for each cell.
In the context of EF-UQ, we demonstrate coarse projective
integration, which accelerates temporal simulation of the cell
population dynamics, and coarse fixed-point computation combined
with Poincar\'e return maps to efficiently converge
on limit cycle solutions, corresponding to synchronous population
oscillations.
Limits of the applicability of the procedure are discussed,
including an extension of the basic methodology to handle loss of
synchronization when isolated outlier cells ``detach" from the main
coherent population and develop individual oscillatory
characteristics.

\section*{A MECHANISTIC SYNCHRONIZATION MODEL FOR YEAST GLYCOLYTIC OSCILLATIONS}

The yeast {\it Saccharomyces cerevisiae\/} exhibits
autonomous oscillations with a period of approximately one minute
when grown under anaerobic
conditions~\citep{aon:cortassa:westerhoff:vandam:1992,
dano:sorensen:hynee:1999, das:busse:1991}.
Similar oscillations have been observed in other yeast
strains~\citep{betz:chance:1965, das:busse:1985} as well as
algae~\citep{kreuzberg:martin:1984}, muscle~\citep{tornheim:1988},
heart~\citep{chance:williamson:jamieson:schoener:1965} and
tumor~\citep{ibsen:schiller:1967} cells.
Yeast studies suggest that an autocatalytic reaction
involving the glycolytic enzyme phosphofructokinase is the main
cause of oscillations at the single cell level. Therefore, the
observed limit cycle behavior has been termed glycolytic
oscillations.
Additional experimental work has focused on characterizing the
intercellular mechanisms involved in the synchronization of
individual yeast cell oscillations~\citep{ghosh:chance:pye:1971}.
Typically, oscillations at the cell population level are observed by
continuous monitoring of the average intracellular NADH
concentration using fluorometry.
Experiments with {\it Saccharomyces cerevisiae\/} grown in anaerobic
stirred cuvettes suggest that secreted acetaldehyde is the key
signaling molecule in the synchronization
mechanism~\citep{richard:bakker:teusink:vandam:westerhoff:1996,
richard:diderich:bakker:teusink:vandam:westerhoff:1994}.

A number of simple cell models have been developed to capture the
glycolytic oscillation mechanism in
yeast~\citep{bier:bakker:westerhoff:2000, goldbeter:lefever:1972,
selkov:1975, wolf:heinrich:1997}.
Cell models based on more
detailed descriptions of the glycolytic reaction pathway also have
been proposed~\citep{hynne:01,
wolf:passarge:somsen:snoep:heinrich:westerhoff:2000}.
Small
ensembles of single cell models have been used to investigate the
synchronization phenomenon~\citep{bier:bakker:westerhoff:2000,
wolf:heinrich:1997,
wolf:passarge:somsen:snoep:heinrich:westerhoff:2000}.
In this paper, we use a single cell model of intermediate
complexity~\citep{Wolf:00} to demonstrate our
computational framework for simulating large populations of coupled
biological oscillators.
Our cell ensemble model is based on
an intracellular coupling mechanism involving the transport of
acetaldehyde across the cell
membrane~\citep{henson:muller:reuss:2002, Wolf:00}.

A single cell in the population is described by the following
differential equations:

 \beqa \!\!\!\!\!\!\!\!\! \frac{dS_{1,i}}{dt}
\!\!\! &=& \!\!\! J_{0,i} - v_{1,i} = J_{0,i}
-k_1S_{1,i}A_{3,i} \left[1+\left(\frac{A_{3,i}}{K_I}\right)^q\right]^{-1} \\
\!\!\!\!\!\!\!\!\! \frac{dS_{2,i}}{dt} \!\!\! &=& \!\!\!
2v_{1,i}-v_{2,i}-v_{6,i} =
2k_1S_{1,i}A_{3,i}\left[1+\left(\frac{A_{3,i}}{K_I}\right)^q\right]^{-1}
- k_2S_{2,i}(N-N_{2,i}) - k_6S_{2,i}N_{2,i} \\
\!\!\!\!\!\!\!\!\! \frac{dS_{3,i}}{dt} \!\!\! &=& \!\!\! v_{2,i}-v_{3,i} = k_2S_{2,i}(N-N_{2,i}) - k_3S_{3,i}(A-A_{3,i}) \\
\!\!\!\!\!\!\!\!\! \frac{dS_{4,i}}{dt} \!\!\! &=& \!\!\! v_{3,i}-v_{4,i}-J_i = k_3S_{3,i}(A-A_{3,i}) - k_4S_{4,i}N_{2,i} - J_i \label{eq:s4ode} \\
\!\!\!\!\!\!\!\!\! \frac{dN_{2,i}}{dt} \!\!\! &=& \!\!\!
v_{2,i}-v_{4,i}-v_{6,i} = k_2S_{2,i}(N-N_{2,i}) -
k_4S_{4,i}N_{2,i}
- k_6S_{2,i}N_{2,i} \\
\!\!\!\!\!\!\!\!\! \frac{dA_{3,i}}{dt} \!\!\! &=& \!\!\!
-2v_{1,i}+2v_{3,i}-v_{5,i} =
-2k_1S_{1,i}A_{3,i}\left[1+\left(\frac{A_{3,i}}{K_I}\right)^q\right]^{-1}
+2k_3S_{3,i}(A-A_{3,i})-k_5A_{3,i} \eeqa

\noindent where the index $i$ denotes the cell.
The pathway model
accounts for glucose flux into the cell ($J_0$), metabolism of
glucose to produce intracellular glycerol, ethanol and a combined
acetaldehyde/pyruvate pool (hereafter called acetaldehyde),
acetaldehyde flux out of the cell ($J$), and degradation of
extracellular acetaldehyde by cyanide.
Glycolytic intermediates modeled are intracellular glucose
($S_1$), the glyceraldehyde-3-phosphate/ dihydroxyacetonephosphate
pool ($S_2$), 1,3-bisphosphoglycerate ($S_3$) and intracellular
acetaldehyde ($S_4$).
Each co-metabolite pair is assumed to be conserved with $A$
and $N$ denoting the constant concentrations of the ADP/ATP and
NAD$^+$/NADH pools, respectively.
Therefore, only the NADH
($N_2$) and ATP ($A_3$) concentrations are treated as independent
variables.
The intracellular reaction rates $v_2$--$v_6$ depend
linearly on the metabolite and co-metabolite involved in each
reaction, while the acetaldehyde degradation rate, denoted $v_7$,
depends linearly on the extracellular acetaldehyde concentration.
Individual cell oscillations are attributable to the
nonlinear term in the reaction rate $v_1$ that accounts for ATP
inhibition.

The net flux of acetaldehyde from the $i$-th cell into the
extracellular environment is modeled as $J_i =
\kappa(S_{4,i}-S_{4,ex})$ where $S_{4,ex}$ is the extracellular
acetaldehyde concentration and $\kappa$ is a coupling parameter
related to the cell permeability.
A mass balance on extracellular
acetaldehyde is derived under the assumption that the volume
fraction of cells relative to the total medium volume ($\varphi$)
remains constant as the total number of cells $M$ is varied:

\vspace{-0.15in} \beq \frac{dS_{4,ex}}{dt} =
\frac{\varphi}{M}\sum_{i=1}^{M} J_i - v_7 =
\frac{\varphi}{M}\sum_{i=1}^{M} \kappa(S_{4,i}-S_{4,ex}) -
kS_{4,ex} \label{eq:s4exode} \eeq \vspace{-0.15in}

\noindent where $k$ is the kinetic constant of the acetaldehyde
degradation reaction.
The chosen parameter
values~\citep{henson:muller:reuss:2002} produce an asymptotic
solution in which all cells are synchronized regardless of the
cell number.
The total number of differential equations ($n$) in the cell
ensemble model increases linearly with the number of intracellular
metabolites (6) and the number of cells ($M$): $n = 6M+1$.
Unless otherwise stated, the following simulations involve
1000 cells: $n = 6001$.

\section*{APPLICATION OF EF-UQ COMPUTATION TO YEAST GLYCOLYTIC OSCILLATIONS} \label{Results:sec}
\subsection*{Polynomial chaos representation of cell properties}
Our cell ensemble model of yeast population dynamics
consists of a large set of coupled nonlinear ODEs.
The intracellular metabolite concentrations in such problems
are in general random variables; many different sources of
randomness exist, from intrinsic kinetic fluctuations, to population
heterogeneity, to randomness in the initial conditions.
In our particular case we consider the randomness arising from {\em
population heterogeneity}; there is a {\em single} random parameter,
the glucose flux $J_0$: $J_0 = \bar{J}_0 + \sigma_J \xi(\omega)$,
where $\xi$ is normal over the sampling space $\Omega$.
This simple choice of a normally distributed uncertainty is made
for illustration/validation purposes; the procedure
is directly applicable to different distributions of 
uncertainty, as we will discuss below.
As our heterogeneous population evolves, each intracellular
concentration evolves, and so therefore do the corresponding
concentration distributions over the population.
The ``obvious" collective variables for evolving distributions are
the first few moments of the distribution: mean, variance etc.; it
might, at first sight, appear that good coarse-grained observables
of our population state would be these first few moments of the
individual intracellular concentration distributions.
These moments, however, do not take into account {\em correlations}
across the distributions \citep{Moon1:06}; 
in our heterogeneous population it is not
enough to know {\em how many} cells have a certain intracellular
metabolite concentration -- we must also know {\em which}
cells have this concentration: the cells with higher or with lower
intrinsic values of $J_0$ ?

\begin{figure}
\center\epsfig{file=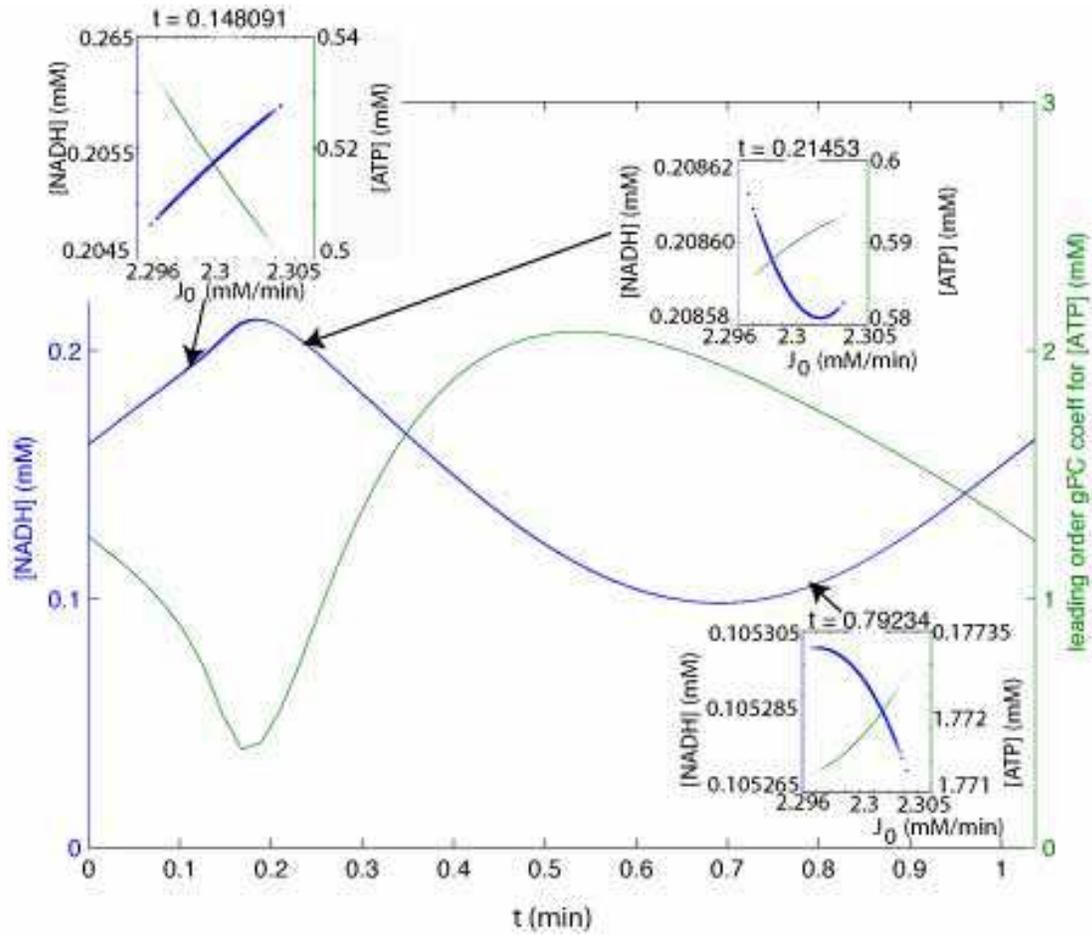,width=\textwidth}
\caption{ 
(color online)
One period of the oscillation of [NADH] for an
ensemble of yeast cells is shown (blue, left axis); 
one oscillation
of the leading order gPC coefficient [ATP] is shown 
(green, right axis). The relationship
of two intracellular concentrations, [NADH] (blue) 
and [ATP] (green), 
with respect to the heterogeneity of the glucose influx
($J_0$) is shown in the inset figures. Note the continuous
dependence of the intracellular concentration on the
parameter $J_0$. 
}
\label{disfulcor:fig}
\end{figure}

We have observed in our simulations (and the same phenomenon has
been documented in different heterogeneous coupled
oscillator contexts \citep{Moon2:06}) that, after a typical
initialization, two distinct phases are observed in the dynamics.
During an initial, {\em relatively fast} phase, strong correlations
develop between the distributions of the various intracellular
concentration values; this is followed by a second, {\em long term}
phase, during which the distributions evolve, but with the
correlations ``locked in".
In effect, these correlations appear to be strongly related to the
heterogeneity of the population - cells with different $J_0$ exhibit
systematically different concentration patterns.
Figure \ref{disfulcor:fig} shows the dependence of concentrations of
NADH and ATP on the population heterogeneity parameter (the random
variable $J_0$) at three instantaneous states. There is clearly a
relation/dependence between the intracellular
metabolite concentrations and the ``identity" of the cell
(the value of $J_0$).

\begin{figure}
\center\epsfig{file=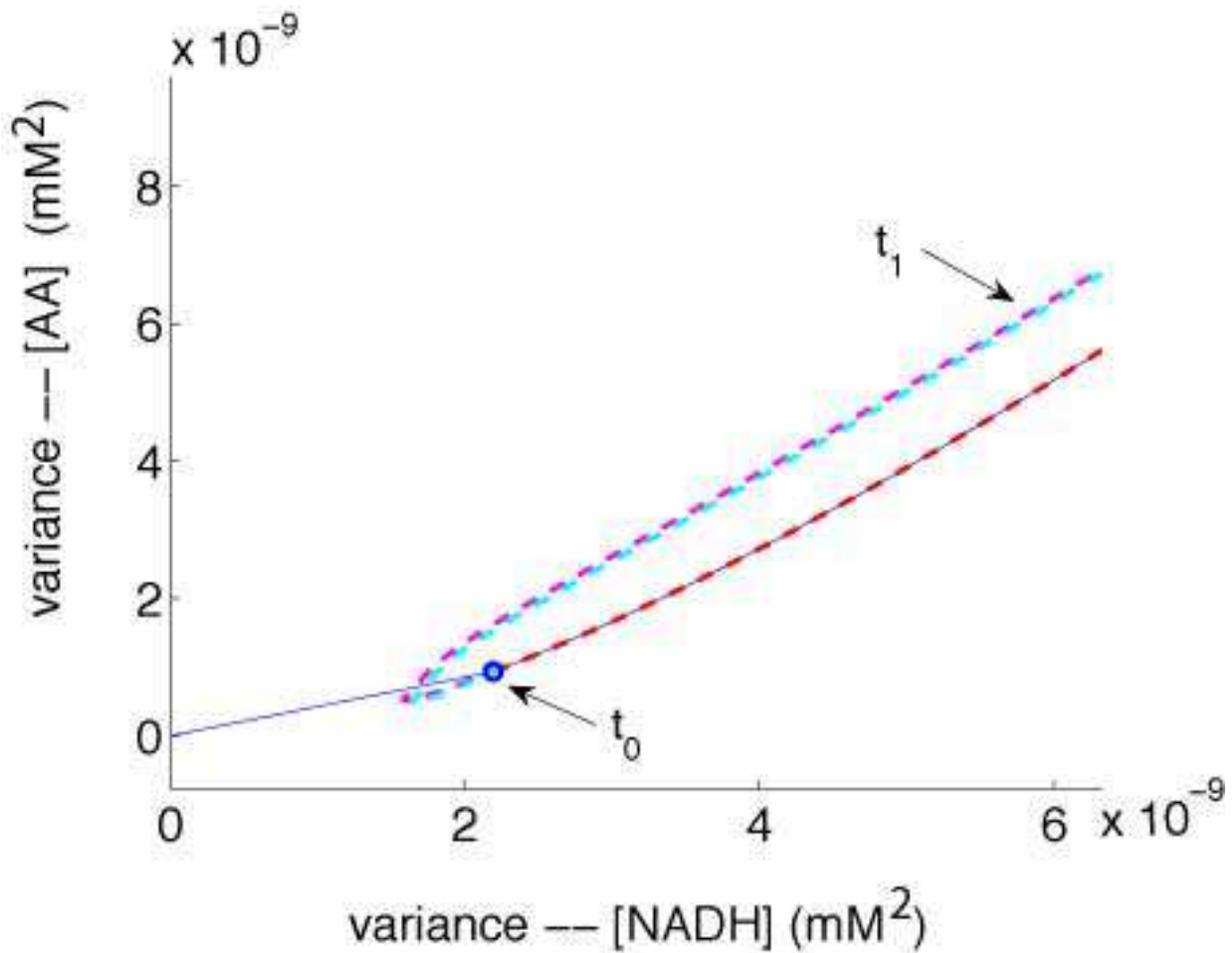,width=\textwidth}
\caption{ 
(color online)
Phase maps for variances of [NADH] and 
[AA] in the initial stage.
Blue curves: variances computed from realizations
located on the fine-level limit cycle.
Red curves: variances computed
from realizations initialized with the fully correlated
lifting from a ``blue" coarse initial conditions. The blue circle indicates
the location of this initial condition.
Magenta and cyan dashed curves: variances
computed from  realizations initialized with two random
liftings from the same coarse initial conditions as above.
Note that blue and red curves almost
coincide with each other. 
}
\label{randomliftingini:fig}
\end{figure}

\begin{figure}
\center\epsfig{file=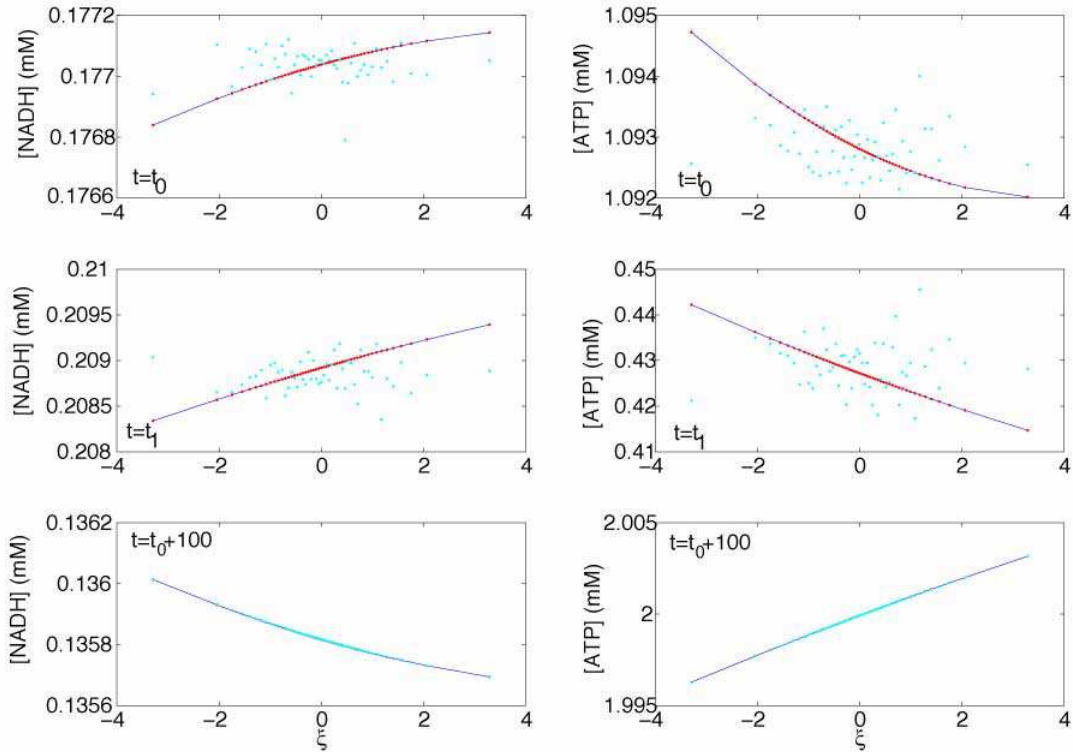,width=\textwidth}
\caption{ 
(color online)
If the
correlations of metabolites within a cell are not
incorporated in the lifting (top: cyan dots at time $t_0$), then
these {\em wrong} correlations remain (middle, $t_1$) until a
sufficiently long amount of time elapses (bottom, $t_1+100$). The
blue line shows the mature, or natural, correlation. The red dots
have been correctly initialized; the cyan points have been
initialized with wrong initial correlations. For consistent
computations, the relationships between the different
metabolites in a cell must be replicated when an ensemble
of cells is constructed (in the {\em lifting} step).
}
\label{disrandliftini:fig}
\end{figure}

If the simulation is continued from the exact state where it was
interrupted, these correlations remain in place (see the red curves
in Figures \ref{randomliftingini:fig} and
\ref{disrandliftini:fig}).
If we now create new, {\em artificial} initial conditions, where
leading moments of the distributions of individual intracellular
concentrations are retained {\em but in which the correlations are
deliberately scrambled} then the system will quickly move away from
the synchronized oscillation in a violent transient (see the cyan
curves in Figures \ref{randomliftingini:fig} and
\ref{disrandliftini:fig}), and will take a long time to return to
it, rebuilding the correlations in the process.
These numerical experiments suggest that strong correlations between
cell heterogeneity and cell behavior get established during initial
stages of the population response, and are then retained in the long
term dynamics.
Based on this observation, and on the functional dependence of
intracellular concentrations on our random variable $\xi$ clearly
apparent in Figure \ref{disfulcor:fig}, we will assume
that all intracellular concentrations across the population can, in
the long-term dynamics, be expressed as (unknown) functions of the
same random variable.

Denoting
$$
{\bm x}(\xi(\omega),t) = (S_1(\xi(\omega),t),S_2(\xi(\omega),t),S_3(\xi(\omega),t),S_4(\xi(\omega),t),N_2(\xi(\omega),t),A_3(\xi(\omega),t))^T,
$$
we will represent ${\bm x}$ in terms of a truncated Polynomial Chaos
expansion of $\xi$
\begin{equation}
  {\bm x}(\xi,t) = \sum_{j=0}^P {\bm x}_c^j(t) \Psi_j(\xi).
\label{singlexiPC:eqn}
\end{equation}
The fine scale state is the $6001$-long vector of dependent
variables (the $6000$ intracellular concentrations plus one
extracellular one) in our detailed set of coupled ODEs.
Our coarse-grained observables are the gPC truncation coefficients,
${\bm x}_c^j(t) =
(x_{1,c}^j,x_{2,c}^j,x_{3,c}^j,x_{4,c}^j,x_{5,c}^j,x_{6,c}^j)^T$.
The {\em lifting} step, the construction of ensemble realizations of
intracellular concentrations (fine-level states) consistent
with a particular set of values of gPC coefficients (coarse-level
observables) is performed through
\begin{equation}
  {\bm x}(\xi_i,t) = \sum_{j=0}^P {\bm x}_c^j(t) \Psi_j(\xi_i), \quad i=1,2,\cdots,M.
\label{ensemblelifting:eqn}
\end{equation}
Here $M(=1000)$ is the total number of cells, the truncation level
$P$ is set to $3$, and $\{ \Psi_j \}$ are orthonormal Hermite
polynomials \citep{Stegun:70} for the case of normally distributed
$\xi$ .
We reiterate that different types of polynomials can be used for
random variables obeying different distributions.
The total number of our {\em coarse} model states is thus
$4\times6+1=25$ (the deterministic extracellular concentration
$S_{4,ex}$ is counted here as a single coarse observable); this is
clearly much less than the number of internal states, $6M+1$, in the
original cell dynamics.

Obtaining the coarse grained observables from a detailed state
constitutes the {\em restriction} step; our restriction protocol
consists of the inner product, Eq. \ref{PCcoefficient:eqn}, which can
be performed in one of two ways: (i) we can discretize the integral
to approximate the inner product, i.e., $<{\bm x}, \Psi_i> = 1/M
\sum_{j=1}^M {\bm x}(\xi_j) \Psi_i(\xi_j)$; (ii) we can perform a
simple least squares fitting to find ${\bm x}_c^j$ such that an $L^2$
norm $||{\bm x}(\xi,t) - \sum_{j=0}^P {\bm x}_c^j(t) \Psi_j(\xi)||$
is minimized.
We used the second implementation to obtain gPC coefficients in this
work.

In what follows, we will demonstrate two distinct ways of using the
EF-UQ methodology in order to accelerate population computations for
our model problem.
We will first demonstrate direct simulation acceleration through
coarse projective integration.
We will then demonstrate the accelerated computation and
coarse-grained stability analysis of synchronized population
oscillations through matrix-free fixed point
computation and eigenvalue approximation; these synchronized limit
cycles will be computed as fixed points of a coarse Poincar\'e map.
By doing so, we demonstrate that our multiscale toolkit
provides a generally applicable and computationally efficient
framework for dynamic simulation and analysis of heterogeneous
populations of cellular oscillators.


\subsection*{Full direct simulation and coarse projective integration}

\begin{figure}
\center\epsfig{file=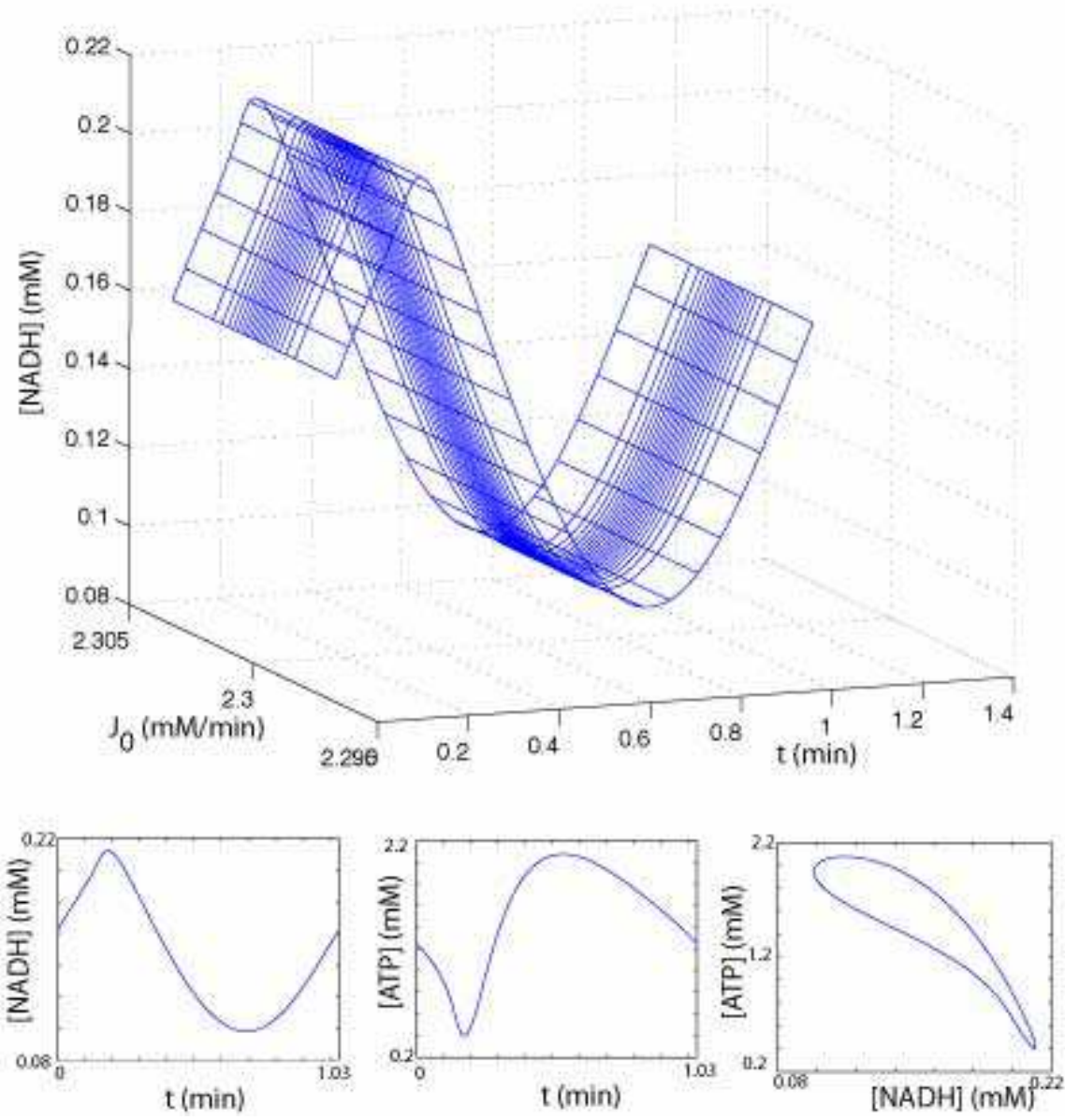,width=\textwidth}
\caption{
Top: The time series of [NADH]
in cells of the ensemble are plotted against time $t$
and the random parameter $J_0$ (glucose influx) 
over one oscillation.
The lines transverse to the time series curve 
are equal-time curves.
Bottom left:
Time series of the leading order gPC coefficients of [NADH]
Bottom right: A projection of the coarse limit cycle onto
the leading order [NADH] and [ATP] gPC coefficient plane.
The distribution of the glucose influx 
parameter $J_0$ has mean
$2.3$ and standard deviation $0.001$.
}\label{fgaussSim}
\end{figure}

The full direct simulation consists of integrating of $6M+1$
differential equations ($M$ is the number of cells, 6
internal states for each cell, and 1 extracellular variable).
The package ODETools is used in Matlab for this simulation, with
variable step-size chosen for relative error tolerance of $1\times 10^{-9}$
and absolute error tolerance of $1\times 10^{-12}$.
Figure \ref{fgaussSim} shows the time
history of such a full simulation of an ensemble of cells over one complete
period for the distribution of parameter $J_0$
having mean $2.3$ and standard deviation $1\times 10^{-3}$.
At discrete moments in time, the values of the full ensemble are
projected to the gPC basis (the full state is {\em restricted} onto
coarse observables), also shown in Figure \ref{fgaussSim}.
At the fine scale, the concentration of each intracellular
metabolite oscillates and, at the end of an oscillation, 
returns exactly to its starting value.
At the coarse, macroscopic level, it is the gPC coefficients
that return to themselves. This is, in effect, saying the
distributions of the species concentrations return to their
initial values.
We now explain the projective integration procedure
and compare its results to the direct simulation.
Starting at $t=t_0$ from a given coarse initial condition (values of
the first few gPC coefficients, the observables), we generate
fine-level model states by the {\it lifting} procedure
Eq. \ref{ensemblelifting:eqn}.
We use the original, full dynamic simulation code to evolve
the fine-level description for an initial time interval $t_{heal}$;
we continue the full direct simulation over three successive time
intervals of time $\delta t$.
At the end of each of these intervals,
the coarse observables are obtained
by {\it restriction}.
The temporal derivatives of the coarse observables
can then be estimated (here for
convenience we use least squares, but maximum-likelihood based
methods are more appropriate \citep{aitsahalia:2002});
these local time derivatives are then
used to {\em project} the values of the coarse variables at a time
$5\delta t$ further in the future.
Using these predicted values we begin the same cycle of brief
healing, detailed simulation observed by restriction, estimation of
local time derivatives and projection of the coarse behavior into
the future.
Depending on the coarse projection
algorithm used in the forward-in-time projection,
more (or fewer) restrictions from the short burst of full
simulation may be needed.
The particular values of $t_{heal}$ and $\delta t$ can also vary from
those used here ($t_{heal} = 5\times 10^{-3}, \delta t = 5\times 10^{-3}$),
and their on-line optimal selection is
problem-dependent.
Figure \ref{fgaussProjInt} demonstrates the 
acceleration of the full
simulation through a coarse projective forward Euler algorithm.
Note that for every $4\delta t$ of full direct simulation, we
project $5\delta t$ into the future -- only $44.44\%$ of the work is
necessary this way.
This type of computational savings enables the simulation of
larger cell ensembles, allowing more accurate reconstructions of
population statistics.
As discussed in \citep{GearA:01} the method can provide significantly
larger savings if a separation of time scales exists in the problem;
for linear problems this can be readily seen as a gap between a few
leading, slow, and the remaining many, fast eigenvalues of the
system.
Different coarse projective algorithms (e.g. projective Runge-Kutta,
or even {\em implicit} projective algorithms) can also be used;
linking them to modern estimation techniques and extending them to
account for adaptive projective step size selection is the subject
of ongoing research (see \citep{GK:2003} as well as \citep{LeeGear:2005}).

\subsection*{Coarse-grained limit cycle computations}

{\bf Limit cycle computation.} Beyond direct simulation, which
asymptotically approaches {\em stable} limit cycles, periodic orbits
are located by solving boundary value problems in time.
In particular, they can be located as fixed points of a {\it
Poincar\'e map} (see, for example, the textbook \citep{Hirsch:04}).
A {\it Poincar\'e Section}, $S$, is a hypersurface (often a
hyperplane) crossing a limit cycle transversely at an (isolated)
point.
The Poincar\'e map, $P: S \rightarrow S$, is a return map defined by
\[P(\xx) = \phi_t(\xx)\in S, \]
for $\xx\in S$, $t$ the smallest positive time for which
$\phi_t(\xx)\in S$, and $\phi_t$ the time $t$ flow map.
Note that it is not necessary to know {\em a priori} the period of
the limit cycle when defining the Poincar\'e map.
Let $P_f$ be the Poincar\'e map for the full simulation and $P_c$
be the Poincar\'e map for the coarse simulation.
The two maps are related by
\[ P_c = R \circ P_f \circ L \]
for $R,L$ the restricting and lifting operators and
$\circ$ is composition.
We use the Newton-Krylov GMRES method to find solutions
of $P_c(\xxc) = \xxc$ \citep{Kelley:95}.
This implementation of Newton iteration does not
explicitly require the computation of the Jacobian of the equation
to be solved (often computed in shooting methods through integration
of the variational equations).
The {\em action} of this Jacobian along sequentially selected
directions (directional derivatives) is estimated from the results
of simulations starting at appropriately chosen nearby initial
conditions.
Since this only requires simulations of the problem, and the linear
equations are solved without ever assembling the relevant Jacobian,
this is a {\em matrix-free} implementation.

Note that the Poincar\'e return map can be constructed, not only
by direct simulation, but also by projective integration.
That is, projective integration can be implemented to accelerate the
computations of the Poincar\'e map itself.
The limit cycle in Figure \ref{fgaussProjInt} was found by
solving this fixed point problem.

{\bf Limit cycle stability.}
For multiscale problems with limit cycles, we can discuss stability at the
fine-scale and also at the coarse-grained level.
We describe here limit cycle stability computations at both levels,
and it is interesting that the results are essentially the same --
the coarse-grained stability
computations reveal the same information
as the (computationally intensive) fine-scale computations.

\begin{figure}
\center\epsfig{file=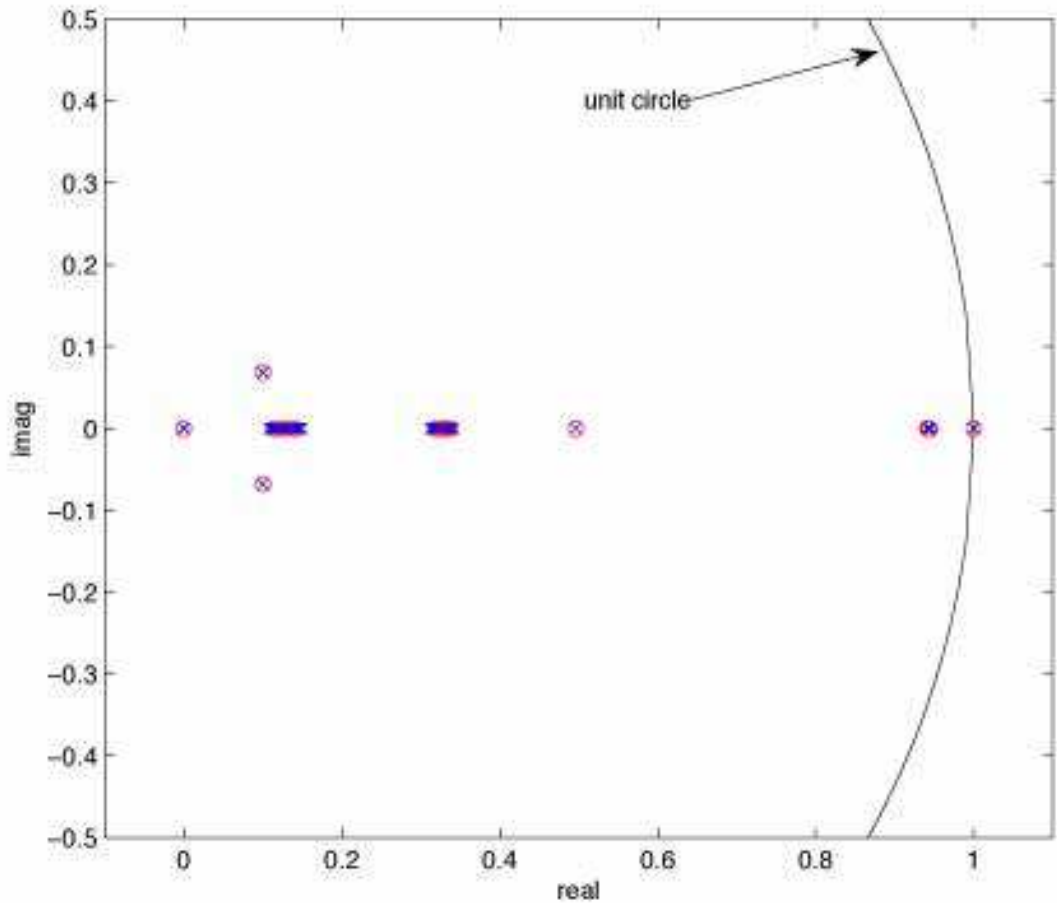,width=\textwidth}
\caption{
(color online)
Leading eigenvalues of $D\phi_t^f$ (blue, $\times$) and
$D\phi_t$ (red, $\circ$). The leading eigenvalues are 
essentially the same, as expected.
the unit circle is shown (black curve); note the
eigenvalues at 1, and all other eigenvalues are inside the unit
circle. analyzing stability of the coarse problem gives
insight into the stability of the full cell ensemble.
}\label{feigens}
\end{figure}

Let $T$ be the period of the limit cycle, and let $\xx_0$ be a point
on the limit cycle.
The time $T$ flow map is
$ \phi_T(\xx_{0}) = x(t; \xx_0)$.
Eigenvalues ({\it Floquet multipliers}) of the matrix $D\phi_T(\xx_0)$
(the Jacobian of $\phi_T$ at
$\xx_0$, the so-called {\em monodromy matrix}) quantify the
linearized stability of the limit cycle.
The time $T$ {\em coarse} flow map is
$\Phi_T = \Phi_T(\xxc)$, see Eq. \ref{eqn:integralCoarseFlow}.
The matrix $D\Phi_T(\xxc)$ can be approximated by finite differences,
\[ \left[  \frac{\Phi_T(\xxc + \ve e_1) - \xxc}{\ve},
\frac{\Phi_T(\xxc + \ve e_2) - \xxc}{\ve}, ... \right] \]
for $\ve << 1$.
The eigenvalues of $D\Phi_T(\xxc)$ are shown in Figure
\ref{feigens}, and the leading ones are listed in Table
\ref{tab:eigenvalues}.
Note that the number $1$ is an eigenvalue (corresponding to time
translational invariance {\em along} the limit cycle at the point
$y_{0}$).
This invariance gives rise to a neutrally stable direction; all
limit cycles possess this neutral eigenvalue at $1$.
The magnitudes of the remaining leading eigenvalues of the limit
cycle are less than 1 (they lie inside the unit circle).
Therefore, the limit cycle is stable.

Eigenvalues of the fine-scale monodromy matrix
reveal the stability of the full limit cycle.
Divided differences could
in principle be used to approximate the full $6001 \times 6001$
Jacobian of the fine scale problem; instead, to obtain this matrix we
integrated the variational equations (see, for example,
the textbook \citep{Hirsch:04}).
Let $\dot \xx = f(\xx)$; then the variational equations along the
solution $x(t,\xx_0)$ are
\[ U'(t,\xx_0) = Df\left(\xx(t,\xx_0)\right) \cdot U(t,\xx_0), \]
where $U(t)$ is a $(6M+1)\times (6M+1)$ matrix and the initial
condition $U(0)$ is the identity matrix.
This gives
\[ D\phi_T^f(\xx_0) = U(T,\xx_0). \]
The ``fine" eigenvalues are shown in Figure \ref{feigens},
and the leading ones are in Table \ref{tab:eigenvalues}.
It can be shown that eigenvalues of the coarse and
fine-scale monodromy matrices
are theoretically the same (if sufficiently many gPC coefficients
are kept), and the coincidence of the eigenvalues
is seen in the table.
Analysis of the fine-scale model considered here is
computationally tractable; for larger problems, however,
a full analysis may not be possible, and the coarse stability analysis
suffices to characterize stability of the fine-scale limit cycles.
For instance, these methods could be profitably applied to
large models of coupled biological oscillators involved in yeast
respiratory oscillations \citep{henson:2004} and mammalian circadian
rhythm generation \citep{to:2006}.

\subsection*{When coarse-graining fails: ``rogue" oscillators}
\begin{figure}
\center\epsfig{file=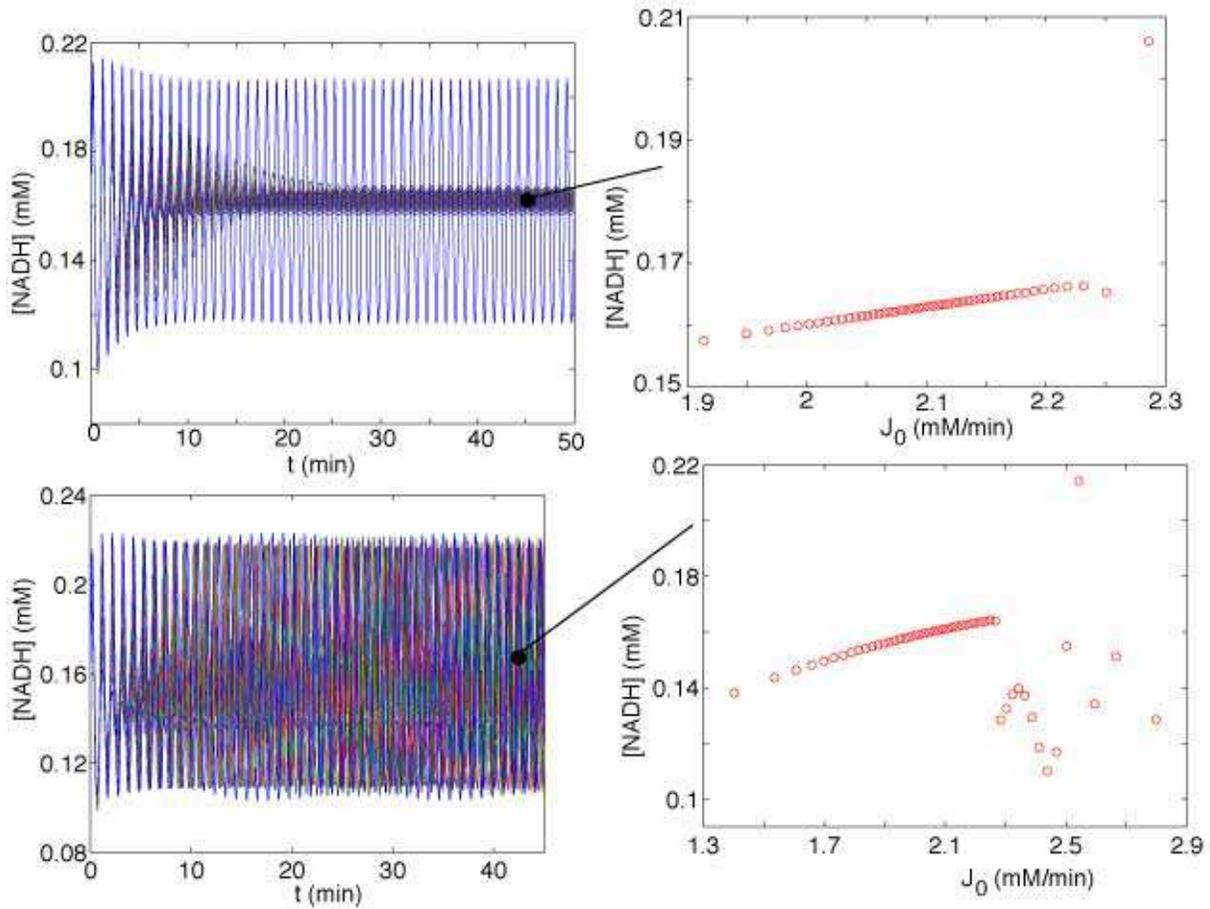,width=\textwidth}
\caption{
(color online)
The discontinuity in the relationship
between [NADH] and the parameter $J_0$ is shown
for distributions with ``rogue" oscillators.
Top:
$J_0$ has a Gaussian distribution with mean and
standard deviation being 2.1 and 0.08, respectively;
bottom: mean $2.1$ and standard deviation $0.3$.
The figures on the left show the time history of [NADH]
for all 50 cells. The figures on the right show [NADH]
as a function of $J_0$ at a snapshot in time
($t = 45$).
Equation-free computations can be applied in these regimes
by using more and different coarse variables: gPC coefficients
for the ``bulk" of the oscillating cells and either
gPC coefficients for the other cells or using all the state
variables of the other cells.
}
\label{discontinuity:fig}
\end{figure}

When the variance of $J_0$ is small, NADH concentrations ($N_2$) in
the cells across the population are narrowly distributed; an overall
synchronized solution for the entire population prevails, and our
coarse-graining methods are successful.
When the variance of $J_0$ increases, however, the amplitudes of the
$N_2$ oscillations across the cell population will separate
significantly.
For certain combinations of mean and variance of $J_0$, one or more
cells will appear to oscillate ``freely" in amplitude.
An illustration of this, using a 50 cell ensemble as the
full direct simulation, and setting the mean and standard deviation
of $J_0$ to 2.1 and 0.08, respectively is seen in Figure
\ref{discontinuity:fig}.
The cause of this single ``free, oscillating" cell can be clearly
rationalized based on the relatively large variance of $J_0$: the
values of  $J_0$ across the population spread {\em beyond}  the
parameter point at which the single cell dynamics undergoes a Hopf
bifurcation.
The strong correlation between individual cell $J_0$ values and
their detailed states (intracellular concentrations) is retained for
{\em most} of the cells, but lost for the outlying ``rogue"
oscillatory cell.
As $\sigma_J$ continues increasing, the strong correlations that
allowed us to use a gPC expansion for the collective behavior fail
even more dramatically - see Figure \ref{discontinuity:fig}, which
is quite representative of dynamics in populations with large
variance of the glucose influx.
We expect that such strong discontinuities in the relation between
heterogeneity and detailed state (reminiscent, at some level, of
Gibbs phenomena) will, in general, be encountered in probabilistic
problems involving strong nonlinearity and bifurcations of the
single oscillator behavior as the heterogeneity parameter(s) is
varied.
Clearly, a few gPC coefficients are no longer good observables of
the collective system state.
The recent literature includes some efforts in resolving such cases
using a wavelet-based chaos expansion \citep{Maitre:04} or piecewise
Polynomial Chaos \citep{Deb:01}; yet the problem remains an open
subject for future research.

Here we will limit ourselves to the case of a {\em single} rogue
oscillator, shown above.
A simple and rational way to tackle this problem is to employ gPC
coefficients to describe the oscillators that are ``clumped
together" with smaller oscillation amplitudes, and use an
additional, {\em different} set of variables to describe
intracellular metabolite concentrations for the single
``freely oscillating" cell.
More specifically, for a total number $M$ of cells which include one
``free" cell, the coarse observables are comprised of
$\bm{x}^0_c,\bm{x}^1_c,\cdots,\bm{x}^P_c$ (obtained by restricting
intracellular concentrations of $M-1$ clumped cells through
the least-square fitting method mentioned earlier), the
extracellular concentration $S_{4,ex}$ and six
intracellular concentrations $S_1,S_2,S_3,S_4,N_2,A_3$
representative of the free cell.
When lifting is now implemented, only intracellular
concentrations of $M-1$ cells are generated from
$\bm{x}^0_c,\bm{x}^1_c,\cdots,\bm{x}^P_c$ through
Eq. \ref{ensemblelifting:eqn}.
The intracellular concentrations  $S_1,S_2,S_3,S_4,N_2,A_3$
for the free cell are part of the coarse-grained description.
Therefore, the full direct simulation is again characterized by
$6M+1$ variables, but the number of observables of the reduced,
coarse-grained problem increases to $4 \times 6 + 1 + 6 = 31$ if the
first four leading order gPC coefficients are retained ($P=3$).

\begin{figure}
\center
\epsfig{file = 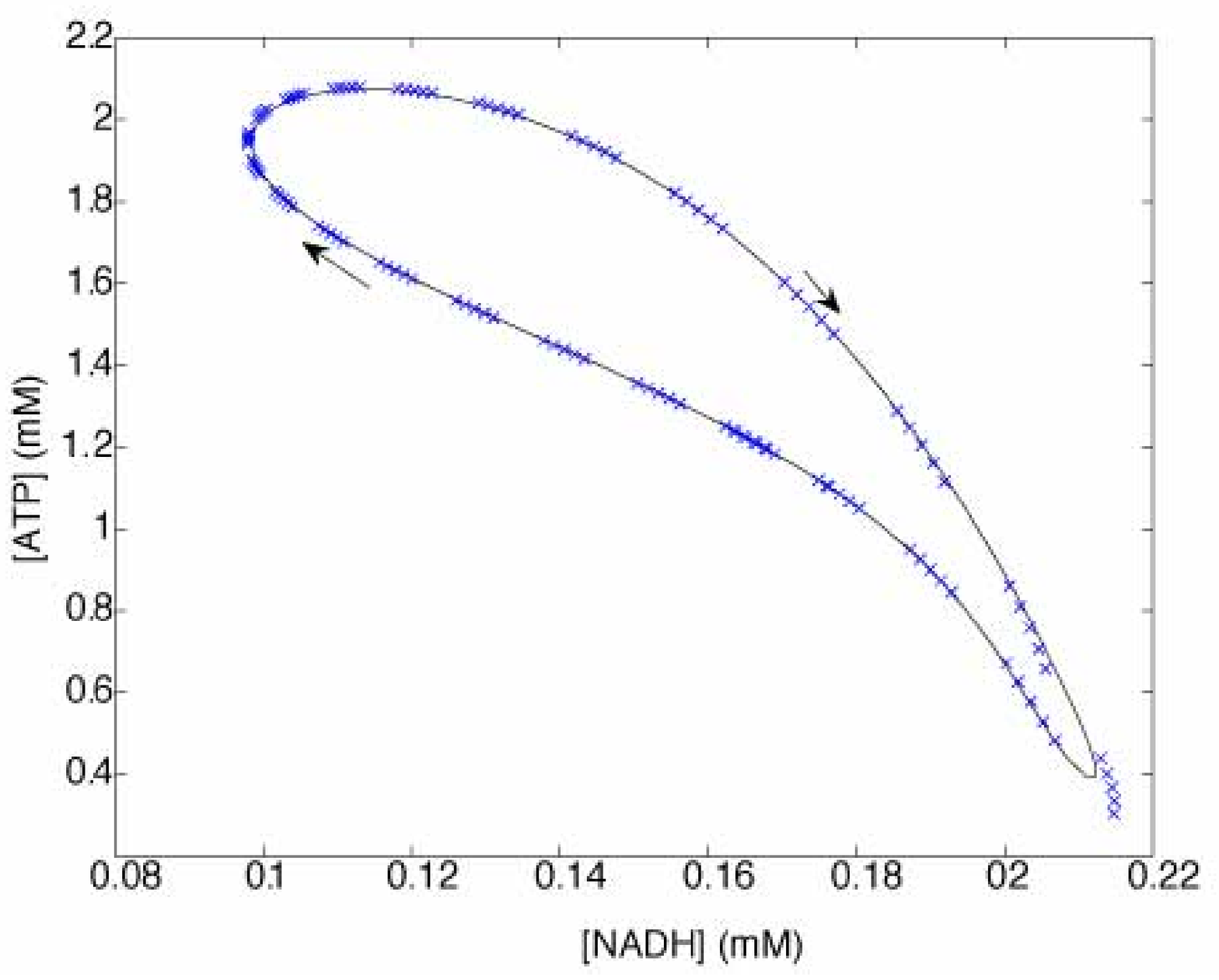,width = 4in}
\epsfig{file=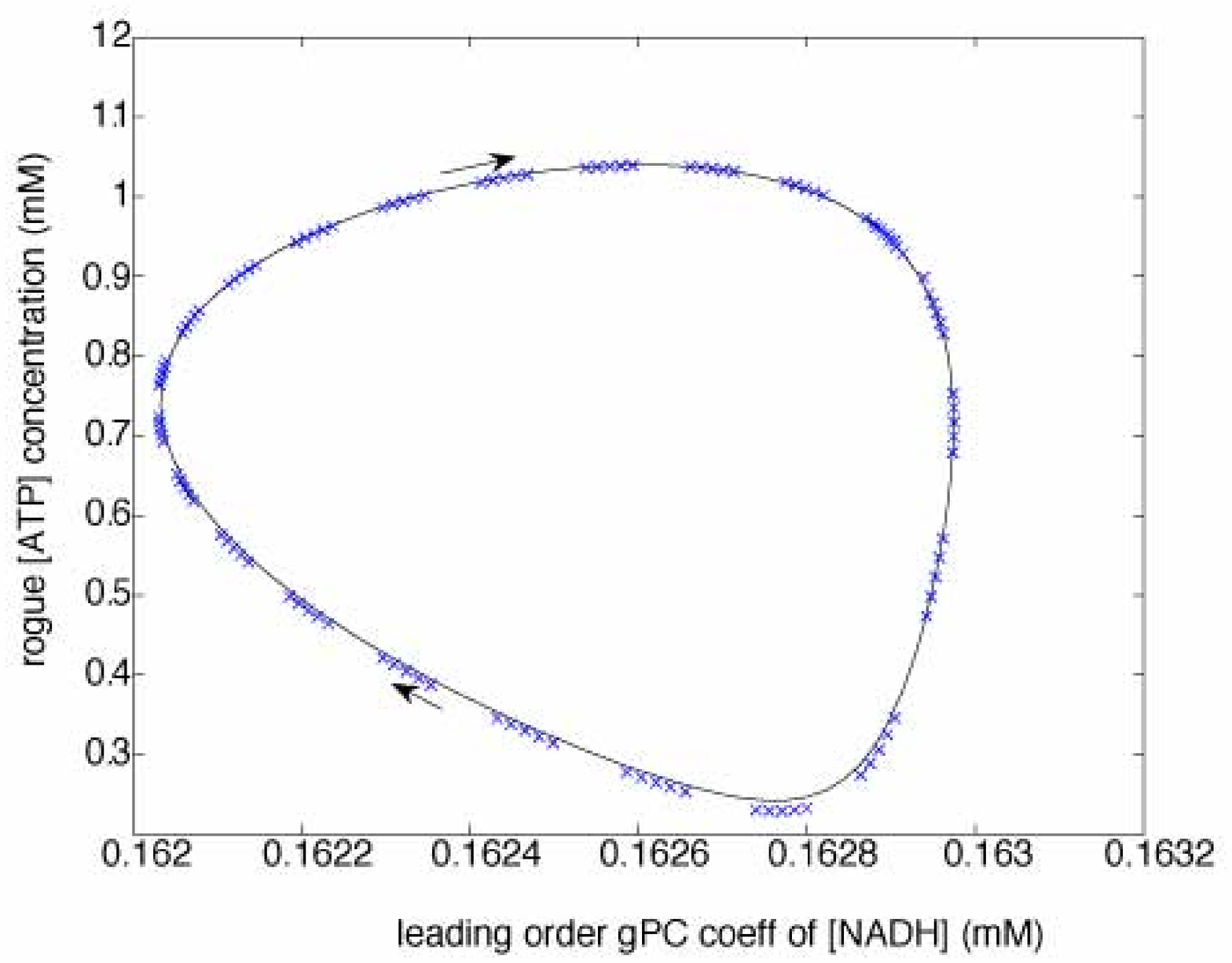,width= 4in}
\caption{ 
Projective integration in coarse variables (blue)
and natural evolution (black). 
Projective integration accelerates
computation of the evolution of a cell population.
Top: An ensemble of size $M=1000$ with the distribution of the
glucose influx parameter having mean $2.3$ 
and standard deviation $0.001$.
Bottom: An ensemble of $50$ cells, 
with the distribution of the glucose
influx parameters having mean $2.1$ 
and standard deviation $0.08$.
In this regime, there is one free cell, 
which oscillates at an amplitude
relatively larger than the others.
For each cell population subplot,
the horizontal axis is the leading order gPC coefficient
for [NADH].
Top: the vertical axis is the leading order gPC coefficent
for [ATP]. 
Bottom: the vertical axis is [ATP] of the
``rogue" cell.
}\label{fgaussProjInt}
\end{figure}

The initial condition on coarse observables for the projective integration
is found by using the equation-free fixed-point algorithm in conjunction
with the Poincar\'e map on the 31-dimensional space.
Coarse projective integration for this new, $31$-dimensional coarse
grained system is illustrated in  Figure \ref{fgaussProjInt}.
Our coarse initial condition is the restriction of a point on the
detailed, fine-scale limit cycle solution; starting there, we use
short bursts of full direct simulation to estimate the time
derivatives of our $31$ coarse observables, and then coarse
projective forward Euler to project their values forward in
time.

The phase portrait of the coarse-grained limit cycle, projected on
the leading order gPC coefficient of the concentration of
NADH of the $M-1$ clumped cells
and the concentration of ATP in the free cell, is shown in
Figure \ref{fgaussProjInt}.
Reasonable numerical agreement with the full direct oscillation
(observed on the same variables) prevails; even in the presence of
one (more generally, of a few) free cell(s), choosing a good set of
coarse-grained variables allows us to accelerate the computations of
the long-term system dynamics.
Of course, this approach requires {\it a priori}
identification of the rogue oscillating cells to construct the
coarse variables.
Clearly more research is needed towards the selection
of good coarse observables for such problems.

\section*{SUMMARY and CONCLUSIONS}
Equation-Free Uncertainty Quantification methods were used in this
paper to accelerate the computer-aided analysis of the dynamics of
{\em heterogeneous} ensembles of coupled biological oscillators; in
particular, coarse-grained computations of synchronized
population-wide limit cycles and their stability was demonstrated
for an ensemble of yeast glycolytic oscillators coupled by
membrane exchange of intracellular acetaldehyde.
The feasibility of the EF UQ in describing certain particular
situations (where one --or a few-- oscillator(s) move freely in
amplitude, distinguished from the ``bulk" of the population) was
also demonstrated.

This paper contains only representative ``proof of
concept" computations.
There is a clear necessity for extensive 
numerical analysis of the
schemes illustrated, including adaptive step-size 
selection, error
estimation and control.
Different restriction schemes need to be devised when the relation
between heterogeneity and behavior becomes nonsmooth or
discontinuous in the ``randomness direction(s)".
Different {\em lifting} schemes \citep{GKKZ} exploiting a
non-explicit separation of time scales may reduce the number of gPC
coefficients required for an accurate reduced description.
Furthermore, a wealth of data-mining techniques currently
under development (and, in particular, the diffusion map approach
\citep{BelkinNiyogi,NadlerLafonCoifmanK}) holds the promise of
extracting low-dimensional parameterizations 
of high-dimensional data
based on graphs constructed on simulation data and the
eigenfunctions of diffusion operators on these graphs.
It would be interesting to explore the performance of such
techniques when simple gPC observables fail, as in the case
of the discontinuities and ``multiple oscillator clumps"  mentioned
above.

\section*{APPENDIX A: SOME BASIC UNCERTAINTY QUANTIFICATION (UQ) AND
EQUATION FREE UNCERTAINTY QUANTIFICATION (EF-UQ) ISSUES}
\label{EFUQ:sec}

\subsection*{On polynomial chaos expansion of random variables and processes} \label{PC:sec}
Orthogonal polynomials of random variables with an arbitrary probability measure
(Gaussian, uniform, Poisson, binomial, ...)
are called generalized Polynomial Chaos (gPC) \citep{XiuB:02}.
Any functional vector, ${\bm x}(\omega)$, of random variables, ${\bm \xi}(\omega)$ ($= (\xi_1(\omega),\cdots,\xi_n(\omega))^T$),
defined over a probability space $(\Omega,{\cal F}, P)$,
can be expressed in terms of a gPC expansion,

\begin{equation}
   {\bm x}(\omega) = \sum_{i=0}^{\infty} {\bm x}_c^i \Psi_i({\bm \xi}(\omega)),
\label{PCexpansioninfty:eqn}
\end{equation}
where $\Psi_i({\bm \xi}(\omega))$ is the $i$th generalized
Polynomial Chaos which
admits the following orthogonality properties,

\begin{equation}
   <\Psi_i,\Psi_j> = \left\{ \begin{array}{ll}
                            <\Psi_i^2>,  &  {\rm if \quad i=j},          \\
                                  0,     &  {\rm if \quad i \neq j},
                             \end{array}
                     \right.
\end{equation}
and ${\bm x}_c^i$ is the corresponding coefficient of $\Psi_i({\bm \xi}(\omega))$, determined by

\begin{equation}
   {\bm x}_c^i = { {<{\bm x}(\omega), \Psi_i({\bm \xi}(\omega))>} \over {<\Psi_i^2>}}.
\label{PCcoefficient:eqn}
\end{equation}
In the above equations, the inner product $<\cdot,\cdot>$ is defined by

\begin{equation}
   <f({\bm \xi}), g({\bm \xi})> = \int_{\Gamma}  f({\bm \xi})  g({\bm \xi}) p ({\bm \xi}) d {\bm \xi}, \quad {\bm \xi} = (\xi_1(\omega),\cdots,\xi_n(\omega))^T,
\end{equation}
where $p ({\bm \xi})$ is the joint probability measure of $ {\bm \xi}$ and $\Gamma$ the support of $p ({\bm \xi})$.

In the case that ${\bm x}$ is a random field or process having the
form ${\bm x}(\omega,s)$ or ${\bm x}(\omega,t)$ where $s$ and $t$
are, respectively, spatial and time coordinates, the projections of
${\bm x}$ onto the Polynomial Chaos, ${\bm x}_c^i$, must admit a
form that depends on these spatial or time coordinates as well.
In our illustrative example the gPC coefficients evolve in (and thus
also depend on) time.

\subsection*{On the stochastic Galerkin method} \label{SGM:sec}
The stochastic Galerkin method aims at quantifying propagation of
uncertainty in dynamical systems.
In this method, solutions of stochastic systems
are first expressed in terms of a finite linear combination of generalized Polynomial Chaos.
The error resulting from the finite-term expansion is then required to be orthogonal
to test functions, which are normally chosen to be the same as the generalized Polynomial Chaos.
A coupled system of equations for the gPC coefficients can thus be
derived and solved \citep{Ghanem:91}.
Probability distributions and
statistical moments of the solutions can be computed from the gPC
coefficients subsequently.

In particular, let the states of a stochastic system be represented
by a vector ${\bm x}(\omega,t): \Omega \times \mathbb{R} \rightarrow
\mathbb{R}^N$, where $\omega$ is an element in the sampling space
$\Omega$.
This system state is governed by the differential equation

\begin{equation}
  { {d{\bm x}} \over {dt} } = {\bm f} ({\bm x}, {\bm \xi}(\omega)), \quad {\bm x}(\omega,0) = {\bm x}_0(\omega).
\label{ODE:eqn}
\end{equation}
The solution of the above equation can be approximated by a finite
expansion in the form of Eq. \ref{PCexpansioninfty:eqn},
\begin{equation}
   {\bm x}(\omega,t) = \sum_{i=0}^P {\bm x}_c^i(t) \Psi_i({\bm \xi}(\omega)).
\label{PCexpansion:eqn}
\end{equation}
By applying a Galerkin projection, equations governing the gPC
coefficients  ${\bm x}_c^i(t)$ are obtained as

\begin{equation}
   { {d {\bm x}_c^i} \over {dt} } = { 1 \over {<\Psi_i^2({\bm \xi})>} } < {\bm f} (\sum_{i=0}^P {\bm x}_c^i(t) \Psi_i({\bm \xi})), \Psi_i({\bm \xi})> , \quad i=0,1,\cdots,P,
\label{coarserODE:eqn}
\end{equation}
with ${\bm x}_c^i(0) =   { {<{\bm x}_0, \Psi_i({\bm \xi})>} \over {<\Psi_i^2({\bm \xi})>}}$.
The above equation can be rewritten as
\begin{equation}
   { {d {\bm X}_c} \over {dt} } = {\bm H}( {\bm X}_c ).
\label{Galerkin:eqn}
\end{equation}
Here ${\bm X}_c=({\bm x}_c^0, {\bm x}_c^1, \cdots,{\bm x}_c^P)^T$ and ${\bm H}=({\bm h}_0, {\bm h}_1, \cdots, {\bm h}_P)^T$, where

$$
    {\bm h}_i({\bm X}_c)= { 1 \over {<\Psi_i^2({\bm \xi})>} } < {\bm f} (\sum_{i=0}^P {\bm x}_c^i(t) \Psi_i({\bm \xi})), \Psi_i({\bm \xi})>, \quad i=0,1,\cdots,P.
$$
If ${{d {\bm X}_c} \over {dt}} = {\bm 0}$ in the long-time limit,
then Eq. \ref{Galerkin:eqn} has a steady state, which can
then be used to recover the probability distribution of the random
steady state of Eq. \ref{ODE:eqn}.
The stochastic Galerkin method can provide an effective reduction of
a stochastic model if a relatively small truncation in Eq.
\ref{PCexpansion:eqn} above is sufficiently accurate.
The corresponding deterministic model will then be considerably
easier to simulate and analyze than large numbers of realizations in a Monte
Carlo simulation of the original dynamics.

\subsection*{On Equation-Free methods and their application in uncertainty quantification} \label{EFUQM:sec}

The basic building block of equation-free methods
\citep{Theodoropoulos:00,Kevrekidis:03,Kevrekidis:04}
is the
{\em coarse time-stepper},
which consists
essentially of three components: {\it lifting}, {\it
micro-simulation}, and {\it restriction}.
{\it Lifting} is a
procedure to transform a coarse-level state to its fine-level
counterpart and {\it restriction} the converse of lifting.
By employing the model states ${\bm x}(t)$ in Eq. \ref{ODE:eqn} as the
fine-level system state vector, and their low-order gPC coefficients
${\bm X}_c$ as the coarse-level observables, Equation-Free methods
can be used to numerically study the behavior of gPC coefficients
without needing closed form ODEs for their evolution,
Eq. \ref{Galerkin:eqn} \citep{Xiu:05}.
The {\it lifting} and {\it restriction} protocols in our context are
then Eq. \ref{PCexpansion:eqn} and
Eq. \ref{PCcoefficient:eqn}, respectively.
The {\it micro-simulation} is just the simulation of the original
large coupled ODE system Eq. \ref{ODE:eqn}.
Assuming that the long-term coarse-grained dynamics in gPC space lie
on a low-dimensional, attracting {\em slow manifold}, one can use
{\em coarse projective integration} to accelerate the computation of
successive coarse-level system states, i.e., low-order gPC
coefficients ${\bm x}_c^i(t_j),i=0,1,\cdots,P,j=0,1,\cdots,M$,
through {\it restriction} of (one or more) short bursts of fine
scale simulation as follows:
Temporal derivatives ${d {\bm x}_c^i} \over {dt}$ at $t_M$ are
estimated by least-squares fitting of the short-time gPC coefficient
evolution ${\bm x}_c^i(t_j),j=M-k,M-k+1,\cdots,M, k<M$.
These gPC coefficients in conjunction with their locally estimated
temporal derivatives are then used to {\em extrapolate} (in effect,
integrate the gPC coefficients numerically in time) over a
relatively large coarse time interval $T$.
For instance, if the coarse forward Euler projective integration is
used, then the predicted gPC coefficients at a later time $t_M+T$
are obtained by

\begin{equation}
   {\bm x}_c^i(t_M+T) \approx {\bm x}_c^i(t_M) +  {{d {\bm x}_c^i} \over {dt}}(t_M) T, \quad i=0,1,\cdots,P. \nonumber
\end{equation}
These {\em projected in time} gPC coefficients ${\bm x}_c^i(t_M+T)$
are {\em lifted} again to the full fine state level, and a new short
burst of micro-simulation is initiated.
The procedure is repeated until a desired time limit is reached.
Issues of time-step selection, estimation and error control are
important, and often ``the devil lies in these details"; discussing
these numerical analysis issues, however, is not the aim of this
brief exposition, and we refer the reader to
\citep{GearA:01,GK:2003,Gear:02,
    Kevrekidis:03}.

If equation Eq. \ref{Galerkin:eqn} possesses a steady state ${\bm
X}_c^s$, then  ${\bm X}_c^s$ must satisfy an integral form of
Eq. \ref{Galerkin:eqn} given by

\begin{equation} \label{eqn:integralCoarseFlow}
  {\bm X}_c^s = {\bm X}_c^s + \int_{t_0}^{t_0+T}  {\bm H}( {\bm X}_c ) dt
\label{fixedpoint:eqn}
\end{equation}
where ${\bm X}_c(t_0) = {\bm X}_c^s$.
The right-hand side of Eq. \ref{fixedpoint:eqn} can be viewed as a
time flow
$\Phi_T: \mathbb{R}^{N \times (1+P)} \times
\mathbb{R} \rightarrow \mathbb{R}^{N \times (1+P)}$.
Therefore, the steady state ${\bm X}_c^s$ is the fixed-point of the
equation
\begin{equation}
  {\bm X}_c = \Phi_T({\bm X}_c).
\end{equation}
When the above equation is not explicitly available, we can use the
coarse time-stepper to approximate the time flow $\Phi_T$.
Newton's method or other iterative algorithms, often in matrix-free
implementations, can be readily employed to compute steady-state gPC
coefficients, which can be used to reconstruct random
stable/unstable steady states of the original system,
Eq. \ref{ODE:eqn}.
It is also possible to compute limit cycles of the gPC coefficients
through Poincar\'e return maps, thus linking EF methods with the
analysis on random limit cycles.

{\bf ACKNOWLEDGEMENTS} 
This work was partially supported by DARPA,
and by an NSF Graduate Research Fellowship to K.B.

\clearpage

\bibliographystyle{biophysj-natbib}

\bibliography{BoldZouKevrekidisHenson}

\clearpage

\begin{table}
\centering
\begin{tabular}{cc}
\hline
\multicolumn{2}{c}{\bf Leading eigenvalues} \\
coarse eigenvalues & full eigenvalues\\
\hline
 1.000004 & 1.000219  \\
 0.942622 &  0.9426798  \\
 0.496275 &  0.4962654  \\
 0.328651 & 0.3273728   \\
$ 0.100267 \pm 0.068410i $ & $0.100250 \pm 0.0683621 i $ \\
 0.126292 & 0.126075 \\
\hline
\end{tabular}
\caption{
Selected, leading eigenvalues from the 
computations of the coarse level
and of the fine level are shown.
The eigenvalue at 1 corresponds 
to a neutrally stable direction
along the limit cycle.
}\label{tab:eigenvalues}
\end{table}

\clearpage

\clearpage

\clearpage

\clearpage

\clearpage

\clearpage

\clearpage

\clearpage

\end{document}